\documentclass[12 pt]{article}

\usepackage{setspace,relsize}

\usepackage{moreverb}

\usepackage[margin=1.3in]{geometry}

\usepackage[pdftex]{lscape}
\usepackage{amscd}
\usepackage{amsmath}
\usepackage{graphicx}
\usepackage{amsfonts}
\usepackage{amssymb}
\usepackage{setspace}
\usepackage{natbib}
\usepackage{mathrsfs}
\usepackage{multicol}
\usepackage{multirow}
\usepackage{pifont}
\usepackage{enumerate}
\usepackage{epic}
\usepackage{color}
\usepackage{booktabs}
\usepackage{authblk}
\usepackage{epstopdf}
\usepackage{subfig} 
\usepackage{newfloat}  
 
\DeclareMathOperator*{\argmin}{arg\,min}
\DeclareMathOperator*{\argmax}{arg\,max} 

\title{\large Automatic Region-wise Spatially Varying Coefficient Regression Model: an Application to National Cardiovascular Disease Mortality and Air Pollution Association Study}

\author{\small Shuo Chen$^1$\thanks{Correspondence to: shuochen@umd.edu}  , Chengsheng Jiang $^2$\thanks{joint first author}  ,  and Lance Waller $^3$ \\
\small $^1$ Department of Epidemiology and Biostatistics, University of Maryland, College Park, MD 20742, USA \\
 
\small $^2$ Maryland Institute for Applied Environmental Health, University of Maryland, College Park, MD 20742, USA

\small $^3$ Department of Biostatistics and Bioinformatics, Emory University, Atlanta, GA 30322
}

\date{}

\begin{document}
\doublespacing \maketitle
\begin{abstract}
Motivated by analyzing a national data base of annual air pollution and cardiovascular disease mortality rate for 3100 counties in the U.S. (areal data), we develop a novel statistical framework to automatically detect spatially varying region-wise associations between air pollution exposures and health outcomes. 
The automatic region-wise spatially varying coefficient  model consists three parts: we first compute the similarity matrix between the exposure-health outcome associations of all spatial units, then segment the whole map into a set of disjoint regions based on the adjacency matrix with constraints that all spatial units within a region are contiguous and have similar association, and lastly estimate the region specific associations between exposure and health outcome. We implement the framework by using regression and spectral graph techniques. We develop  goodness of fit criteria for model assessment and model selection. 
The simulation study confirms the satisfactory performance of our model.  We further employ  our method to investigate the association between airborne particulate matter smaller than 2.5 microns (PM 2.5) and cardiovascular disease mortality. The results successfully identify regions with distinct associations between the mortality rate and PM 2.5 that may provide insightful guidance for environmental health research. 
  

\end{abstract}

\newpage

%
%
%

\section{Introduction}
The recent environmental health research has revealed that the associations between air pollution exposures and health outcomes vary spatially because the local environmental factors including topography, climate, and air pollutant constituents are heterogeneous across a broad area (\citealp{Bell14}; \citealp{ChungY}; \citealp{Garcia}).  However, the current available region definition (e.g. the 48 states of the continental U.S.) can not ensure homogeneous intra-region exposure-health-outcome association (EHA). Therefore, it is desirable to identify a set of disjoint regions exhibiting similar within-region EHAs and distinct between-region EHAs in a data-driven fashion (automatic rather than predefined region map).  The statistical inferences based on the automatic detected regions (e.g. EHA regression analysis) can provide important guidance for environmental health research. 

Motivated by analyzing a data set from the national data base of annual air pollution exposures and cardiovascular disease mortality rates for 3100 counties in the U.S. during the 2000s (more details are provided in section 4), we aim to estimate the spatially varying associations between air pollution exposures and health outcomes across different regions. The region level inferences may reveal local environmental changes and latent confounders that could influence local population health. However, most current spatial statistical models (for areal data analysis) are limited for this purpose because few of them allow data-driven allocation of counties into contiguous regions (that different regions demonstrate distinct health risks). In our analysis, we use county as the basic spatial unit because the health outcome data is aggregated at the county level, yet the proposed approach could be applied for analysis with any basic spatial unit (e.g. zip code).

The most popular modeling strategy for areal spatial data has been through
the conditionally autoregressive (CAR) distribution and its variants.(\citealp{Besag74}; \citealp{Besag91}; \citealp{Gelfand03a}; \citealp{Banerjee};  \citealp{Waller04}; \citealp{Cressie}). In disease mapping, a random effect model is often employed to link the disease rate with exposures, and the random residuals and random slopes are used to account for spatial dependence via CAR or multivariate CAR (MCAR) priors (\citealp{Gelfand03b}; \citealp{Waller07}; \citealp{Wheeler08}; \citealp{Banerjee08}).
Although the random residuals and slopes could improve  goodness-of-fit by explaining more proportion  of variance, the inferences on the regression coefficients (main effects) are still limited to one value for the whole nation(\citealp{Waller04}). The spatial dependency information is  incorporated into statistical modeling by using a spatial adjacency matrix $W$ where $W_{ij}=1$ if $i$ and $j$ are adjacent neighbors  and  otherwise $W_{ij}=0$. It has been recognized that the spatial adjacency may not necessarily lead to similar random effects because health outcomes related factors such as landscape types (e.g. urban area and forest) or sociodemographic factors (e.g. income levels) could be distinct between neighbors.   \citealp{Lu05} propose a Bayesian hierarchical modeling framework  to detect county  boundaries  by using boundary likelihood value (BLV) based on the posterior distributions, yet the detected boundary segments are often disconnected (\citealp{Ma07}). $L_1$ norm CAR prior has also been applied to account for abrupt changes between neighbors (\citealp{Best}; \citealp{Hodges}).     \citealp{Ma10} further include the random effect for the boundary edges with Ising priors and identify more connected boundary segments. Although the boundaries can provide information of adjacent but distinct neighbors, they may not allow to define distinct regions due to the discontinuity of boundary edges. It is also worth to note that the boundaries are often defined based on the outcomes (or residuals) rather than the disease-exposure associations (regression coefficients). 

To fill this gap, we present an automatic region-wise spatially varying coefficient method to recognize and estimate the spatially varying associations between air pollution exposures and health outcomes in automatically detected regions for environmental health data, and we name it as region-wise automatic regression (RAR) model.
Rather than focusing on modeling the spatial random effects, the RAR model aims to parcellate the whole spatial space into a set of disjoint regions with distinct associations, and then to estimate the association for each disjoint region. We implement the RAR model in three steps: first, we assess the initial difference of associations by examining each spatial unit's impact on the overall regression coefficient based on $dfbeta$;  second, based on the initial difference of the associations, we cluster the all spatial units into a set of spatially contiguous regions by using image segmentation technique; last, we estimate the associations for each region, and we account for the within region spatial correlation of the residuals. We also develop a likelihood based optimization strategy for parameter selection. Our main contribution is first to provide a statistical model to identify the data-driven definition of regions exhibiting differential regression coefficients that may yield significant public health impact. 

The rest of this paper is organized as follows. In Section 2, we describe RAR model and its three step estimation procedure. In Section 3, we conduct simulation studies with the known truth to examine the performance of the RAR model and to compare it with existing methods. In section 4, we apply the proposed method to a environmental health dataset on investigating the association between PM 2.5 and cardiovascular disease mortality rate in the U.S.. We conclude the paper with discussions in section 5.

\section{Methods}
\numberwithin{equation}{section}

We use a graph model $G=\{V,E\}$ to denote the spatial data, where the vertex set $V$ represents all spatial units (e.g. counties) in the space/map and the edge set $E$ delineates the similarity between the vertices. For RAR model, $e_{ij} \in E$ reflects the similarity between the EHA of spatial units $i$ and $j$ ($i,j \in V$ and $|V|=S$). 

In the following subsections, we introduce the three steps of RAR model: 1) to assess the spatial association affinity between spatial units ($e_{ij}$); 2)
to parcellate graph $G$  into $K$ regions that $G= \cup_{k=1}^K G_k$ and $G_k\cap G_{k'}=\emptyset$,  and the spatial units within one region $G_k$ exhibit coherent coefficients; and 3) to estimate the region-specific EHA. For notational simplicity, we only consider cross-sectional study modeling though it is straightforward to extend the RAR method to longitudinal studies.

\subsection{Step 1: The adjacency matrix $ E =\{e_{ij}\}$}
To assess the association between health outcomes and covariates, a Poisson regression model is often used: 
\begin{align}
Y_i|\mu_i & \sim \mbox{Poisson}(\mu_i n_i),  \mbox{ for all } i=1, \cdots, S  \notag\\
log(\mu_i) & =  \beta  \mbox{PM2.5}_i + \boldsymbol{\eta}\textbf{X}_i  
\label{eq:reg}
\end{align}

where $Y_i$ is  death count and  $n_i$ is age-adjusted expected population count for county $i$, and $\textbf{X}_i$ are covariates of potential confounders besides the air pollution exposure.  If the association between air pollution exposure and health outcome at the location $i$ deviates from the general trend, then the regression  coefficient  excluding location $i$   $ \boldsymbol{\beta}_{(i)}$  will be distinct from the general regression coefficient of all observations $ \boldsymbol{\beta}$.  Therefore, we adopt DFBETAs to measure the EHA deviation for location $i$. We  denote $d_i$ as the deviation of spatial unit $i$: 
\begin{align}
d_i = \hat{\beta}_{(i)}-\hat{\beta} \approx \mbox{DFBETA}_i= \{ (Z^T \Omega Z)^{-1} Z_{i}\Omega_i(1-h_i)^{-\frac{1}{2}}r_{pi} \}_{PM2.5} / \hat{\sigma}(\hat{\beta})
\label{eq:dfbeta}
\end{align}

where $\textbf{Z}$ is the design matrix including both PM2.5 and $X$, $\Omega$ is weight matrix ( $\Omega$ is the identity matrix if no weight is assigned), $h_i$ is the leverage, $r_{pi}$ is the standardized Pearson residual, $\hat{\sigma}(\beta)$ is the  estimated standard deviation of $\beta$ (\citealp{Williams}). Note that \ref{eq:dfbeta} is a one-step approximation to the difference for a generalized linear model (\citealp{Preisser96}). 

Then, the similarity of the associations between two spatially adjacent units $i$ and $j$ are calculated by a distance metric (e.g. Gaussian similarity function):
\begin{align}
e_{ij}=\mbox{exp}(-\frac{(d_i-d_j)^2}{2 \widehat{\sigma_d}^2})\times\mbox{I}(i \sim j)
\label{eq:eij}
\end{align}
where $\widehat{\sigma_d}$ is the standard deviation across all $d_i$, and $\mbox{I}(i \sim j)$ is the indicator function which equals 1 only when $i$ and $j$ are  spatially adjacent. In addition, if the natural cubic splines are used to fit the nonlinear trend of the air pollution exposure (\citealp{Dominici02}), then $e_{ij}=exp(-\frac{1}{2} (\textbf{d}_i-\textbf{d}_j)^T \widehat{\Sigma}^{-1}(\textbf{d}_i-\textbf{d}_j))$. Thus, the range of $e_{ij}$ is from 0 to 1. The similarity matrix $E$ is a $S \times S$ matrix, which is equal to $E^0\circ W$ ($E^0=\{ \mbox{exp}(-\frac{(d_i-d_j)^2}{2 \widehat{\sigma_d}^2})\}$ and $\circ$ is Hadamard product ).  Thus, matrix $E$ fuses information of PM2.5 exposure (along with other covariates) and health outcomes in $E^0$ with spatial adjacency information $W$.

\subsection{Step 2: Spectral graph theory based automatic region detection}
The goal of step two is to identify the disjoint and contiguous regions ($G= \cup_{k=1}^K G_k$) such that the EHA of the spatial units are homogeneous within each $G_k$ but distinct between $G_k$ and $G_{k'}$ ($k\neq k'$). The contiguity requires that $\forall i$ vertex in $G_k$, there exists at least one vertex $j$ which is connected to $i$. Then the region detection becomes a graph segmentation problem based on the similarity matrix $E$.  We aim to estimate a set of binary segmentation binary parameters $\Delta=\{ \delta_{ij} \}$ and $ \delta_{ij}= 0 \mbox{ or }1$ that parcellate $G$ into $\{ G_k \}$. Therefore, the segmentation model aims to estimate the latent binary parameters: $\widehat{\delta}_{ij}|e_{ij}, N(e_{ij})$ with the contiguity constraint, 
where $N(e_{ij})$ is the neighborhood of edge $e_{ij}$. One way to circumvent this is to identify the $\widehat{\delta}_{ij}=0$ for those $e_{ij}>0$ in $E$ (which is analogous to ``cutting" edges in the graph) with the optimization that: 

\begin{align}
&\sum_{k=1}^K \frac{\mbox{cut}(G_k, \overline{G_k})}{\mbox{vol}(G_k)} \notag \\ 
& \mbox{cut}(G_k, \overline{G_k})=\sum_{i \in G_k, j \in \overline{G}_k}e_{ij} 
\label{eq:cut}
\end{align}
 where $K$ is the number of disjoint sets and $ \mbox{cut}(G_k, \overline{G}_k)$ is the sum of a set of edge weights  (with $e_{ij}>0$ and $\widehat{\delta}_{ij}=0$) to isolate the subgraph $G_k$ from $G$. However, even for a planary  graph G and $K=2$ parcellation, the optimization is NP complex. To solve the objective function in \ref{eq:cut}, a two step relaxation procedure is often used. The first step is a continuous relaxation:
 
\begin{align}
& \mbox{minimize } \sum _{l=1} ^K f'_l L f_l=\frac{1}{2}\sum _{l=1} ^K \sum_{i \neq j}e_{ij}\Vert  \frac{f _l(i)}{D^{1/2}_{e(i,i)}}  -\frac{f _l(j)}{D^{1/2}_{e(j,j)}} \Vert ^2 \notag \\ 
& L=I-D_E ^{-1}E
\label{eq:relaxation1}
\end{align}

where L is the normalized graph Laplacian matrix with $D_E=Diag(e_{i+})$ ($e_{i+}=\sum_j e_{ij}$), and $f=\{f_1, \cdots f_K\}$ is a $N \times K$ coordinate matrix (all entries are continuous) that places the close nodes (based on the adjacency matrix $E$) near to each other in $\mathbb{R}^K$, and $f_l$ ($l=1 \cdots K$,  $N \times 1$) is the $l$th vector  (\citealp{Chung}). By Rayleigh quotient, $f$ are the first $K$ eigen-vectors of spectral decomposition of $L$ (with ascending order of eigen-values) (\citealp{Von}). In addition, the unnormalized graph Laplacian matrix $L=D_E-E$ is also often used (\citealp{Von}). From the spectral graph theory point of view, the Bayesian CAR model updates the random effect ($\boldsymbol{\phi}$) posterior sampling by using the heuristic to increase the fusion of likelihood and the objective function in \ref{eq:relaxation1} which is equivalent to minimizing  unnormalized  graph Laplacian matrix ($L=D_E-E$). Therefore, the CAR prior can be considered as a penalty (restriction) term that aligns with the spectral clustering objective function.  

The second step of the automatic region detection procedure is discretization relaxation that produces  $f^d$ which is a $N \times K$  binary matrix with all entries either 0 or 1 and $f^{d}  \textbf{1}_{K \times 1} = \textbf{1}_{N \times 1}$. Then, objective function becomes

\begin{align}
 \mbox{minimize } \sum _{l=1} ^K f^{d'}_l L f^d_l=\frac{1}{2}\sum _{l=1} ^K \sum_{i \neq j}e_{ij}\Vert \frac{f^d_l(i)}{D^{1/2}_{e(i,i)}}  -\frac{f^d_l(j)}{D^{1/2}_{e(j,j)}} \Vert ^2,   
 \label{eq:relaxation2}
\end{align} 

which is equivalent to \ref{eq:cut}. The second step optimization aims to calibrates the $f^d$ coordinate matrix with reference to $f$.  To implement this optimization step, \citealp{Ng} and \citealp{shi} apply K-means clustering algorithms for the distretization relaxation. However, the K-means clustering algorithm results may vary due to different random initialization and yield unstable results. \citealp{shi03} develop multiclass spectral clustering algorithm which is robust to random initialization and nearly global-optimal. The optimization yields results of the automatic region detection, and based on $f^d$ all spatial units are categorized to $K$ class with contiguity. In addition, $\widehat{\delta_{ij}}$ can be obtained from the resulting $f^d$, and  under mild regularity condition the spectral clustering based region segmentation estimator is consistent that is $\widehat{\delta_{ij}} \rightarrow \delta_{ij}$ (\citealp{Von08}, \citealp{Lei14}). We briefly summarize the algorithm in Appendix and refer the readers to the original paper for the detailed optimization algorithm.  
 
 As a comparison, the popular Bayesian CAR model  leverages the second line of formula \ref{eq:reg}   $$ log(\mu_i)   = \beta  \mbox{PM2.5}_i + \boldsymbol{\eta}\textbf{X}_i + \phi_i,$$
and imposes areal random intercept $ \boldsymbol{\phi}=(\phi_1,..., \phi_S)$  linked with the spatial adjacency matrix $W$ by letting $\boldsymbol{\phi} \propto exp(-\frac{\tau_{\phi}}{2} \boldsymbol{\phi}'(D-\rho W)\boldsymbol{\phi})$. $\tau_{\phi}$ is a positive scale parameter, $W$ is the spatial adjacency matrix introduced in section 1, $D=Diag(w_{i+})$ ($w_{i+}=\sum_j w_{ij}$), and $\rho$ is chosen to ensure $(D-\rho W)$ non-singular. Note that the distribution of $\phi$ is a proper CAR distribution when $\rho \neq 0$.

When implementing the MCMC for a Bayesian CAR model, the chain updating criteria incorporate both the likelihood part  $lik(y|\boldsymbol{\theta})$   and the CAR prior $ exp(-\frac{\tau_{\phi}}{2} \boldsymbol{\phi}'(D-\rho W)\boldsymbol{\phi})$. The parameter update rule in part favors  smaller $\frac{\tau_{\phi}}{2} \boldsymbol{\phi}'(D-\rho W)\boldsymbol{\phi}$ values, i.e. the spatially adjacent $\phi_i$ and $\phi_j$ have similar values.
Interestingly, the prior function is intrinsically linked with the objective function of spectral clustering algorithm which aims to minimize $tr(\boldsymbol{f^d}'(D-\rho W)\boldsymbol{f^d})=\sum_{l=1}^K \boldsymbol{f^d}_l'(D-\rho W)\boldsymbol{f^d}_l$, where $D-\rho W$ is unnormalized graph Laplacian matrix when $\rho=1$ and $\boldsymbol{f^d}$ is discretized $N \times K$ (N nodes and K classes) matrix to represent the cluster membership that $\boldsymbol{f^d} \in \{ 0,1\}$ and $\boldsymbol{f^d}\boldsymbol{1}_K=\boldsymbol{1}_N$(\citealp{Von}).  Using our similarity matrix $E$ as $W$, $tr(\boldsymbol{f^d}'(D-\rho W)\boldsymbol{f^d})$ is smaller when the spatially adjacent units with similar DFBETAS are classified into one spatial cluster. With similar objective functions, $\boldsymbol{f^d}$ could be obtained by discretizing $\boldsymbol{\phi}$ (\citealp{Von}). 
Hence, both the CAR and RAR model incorporate spatial adjacency in a close format of the updating criteria and objective function. Yet, as the random effects (residuals or the random slopes) $\boldsymbol{\phi}$ in the CAR model are continuous, they could not provide region parcellation information to identify  regions with distinct EHA as the RAR model does.


\subsection{Step 3: Association estimation on the automatically detected regions }

Provided with the automatic region parcellation results $G= \cup_{k=1}^K G_k$, we estimate the region/subgraph specific association between air-pollution exposure and health outcomes. The straightforward method is to conduct stratified analysis such that within each $G_k$, the GLM is estimated:   

\begin{align}
y_i &\sim \mbox{Poisson}(\mu_i) , \ \  \forall i \in G_k \notag \\
g(\mu_i)&=\beta^k  \mbox{PM}_i + \boldsymbol{\eta}^k\textbf{X}_i.
\end{align}

Within each $G_k$, we further investigate the spatial dependence of the residuals by using semivariogram or Moran's I statistic. If the residuals of the spatially close units are  dependent with each other, then the spatial autoregressive models such as CAR and SAR  can be applied for regression analysis.  Alternatively, a GLM could be applied to fit all spatial units by using the region indicator as a categorical covariate and adding the interaction terms of the categorical region indicators and the air pollution exposure. 

In light of the law of parsimony, we aim to maximize the likelihood with constraint of model complexity and apply the commonly used model selection criteria BIC to determine the appropriate number of $K$. Particularly, the BIC value is a function of number of regions ($K$),  and lower BIC implies appropriate number of regions.  
   
%

\section{Simulations}
We conduct a set of simulations to demonstrate the performance of RAR and  compare it with conventional spatial statistical methods including SAR and CAR models. 

\begin{figure}[ht!]
\begin{minipage}{.5\linewidth}
\centering
\subfloat[]{\label{rS:a}\includegraphics[scale=.3]{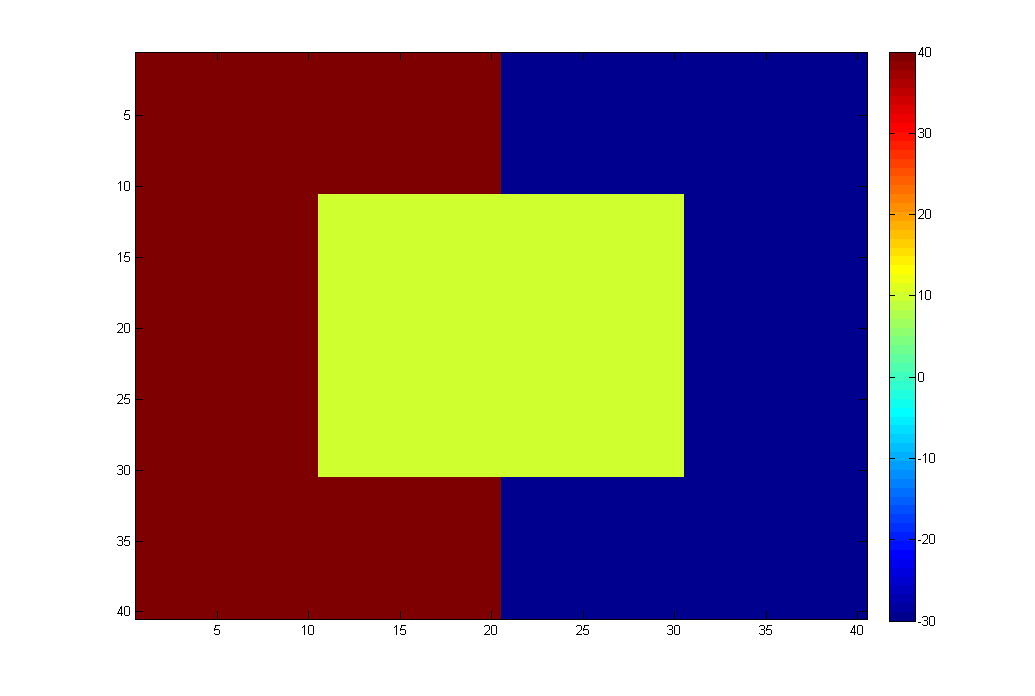}}
\end{minipage}%
\begin{minipage}{.5\linewidth}
\centering
\subfloat[]{\label{rS:b}\includegraphics[scale=.3]{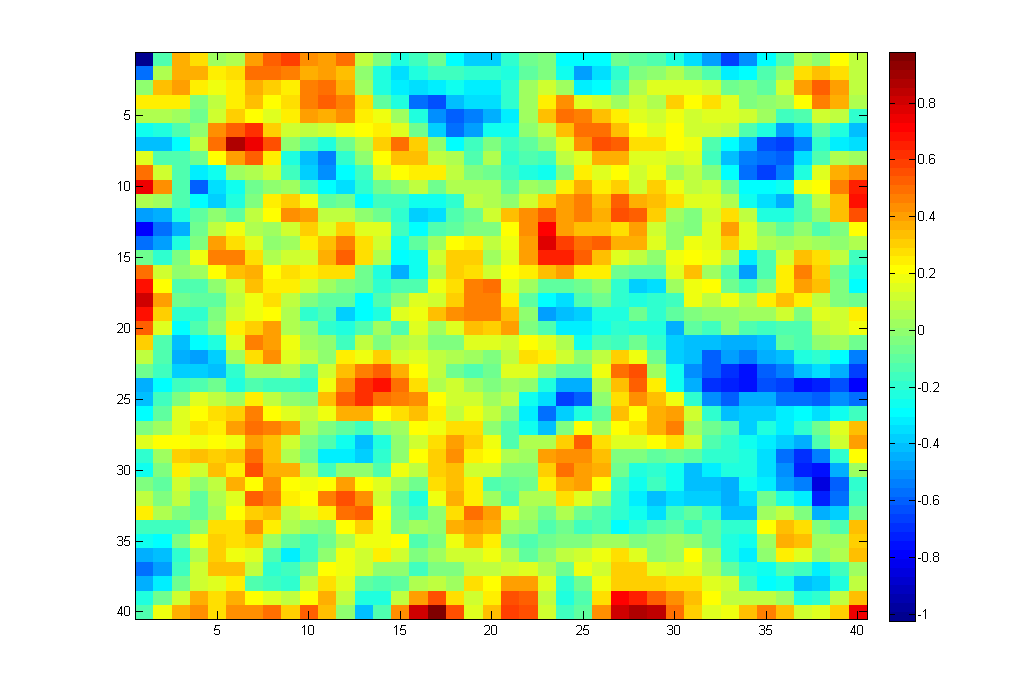}}
\end{minipage}\par\medskip
\vspace{-0.01 in}
\begin{minipage}{.5\linewidth}
\centering
\subfloat[]{\label{rS:b}\includegraphics[scale=.3]{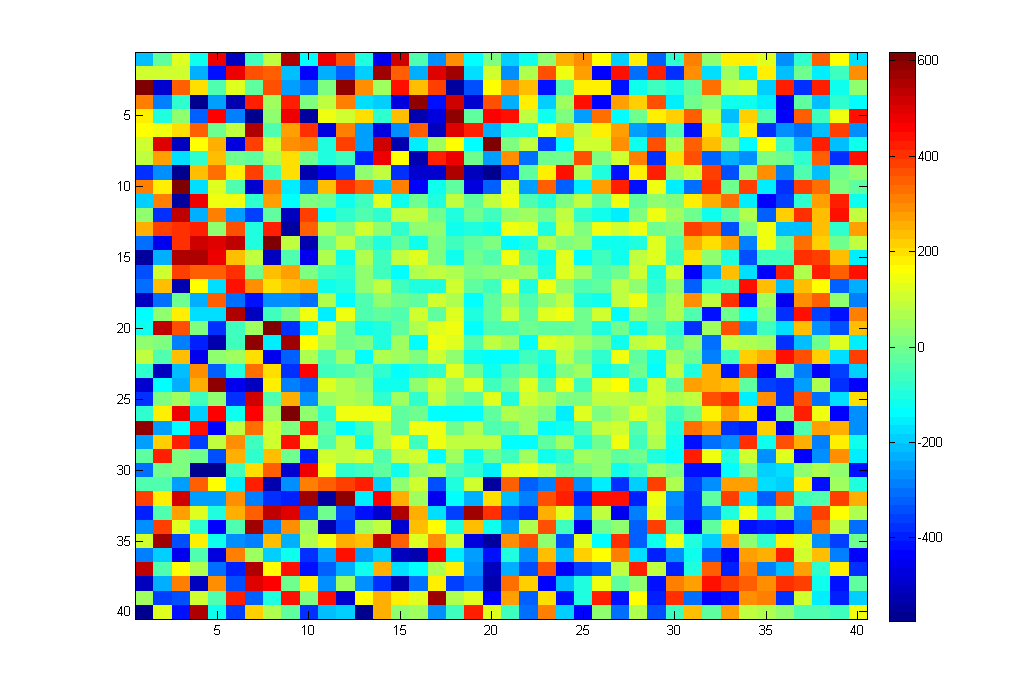}}
\end{minipage}
\begin{minipage}{.5\linewidth}
\centering
\subfloat[]{\label{rS:c}\includegraphics[scale=.47]{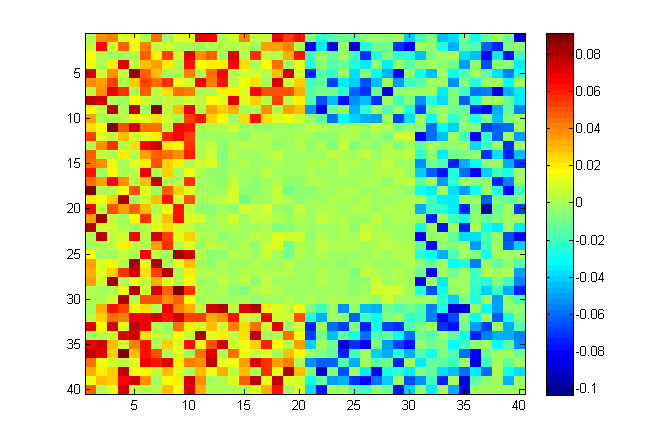}}
\end{minipage}\par\medskip
\vspace{-0.1 in}
\begin{minipage}{.95\linewidth}
\centering
\subfloat[]{\label{rS:a}\includegraphics[scale=.3]{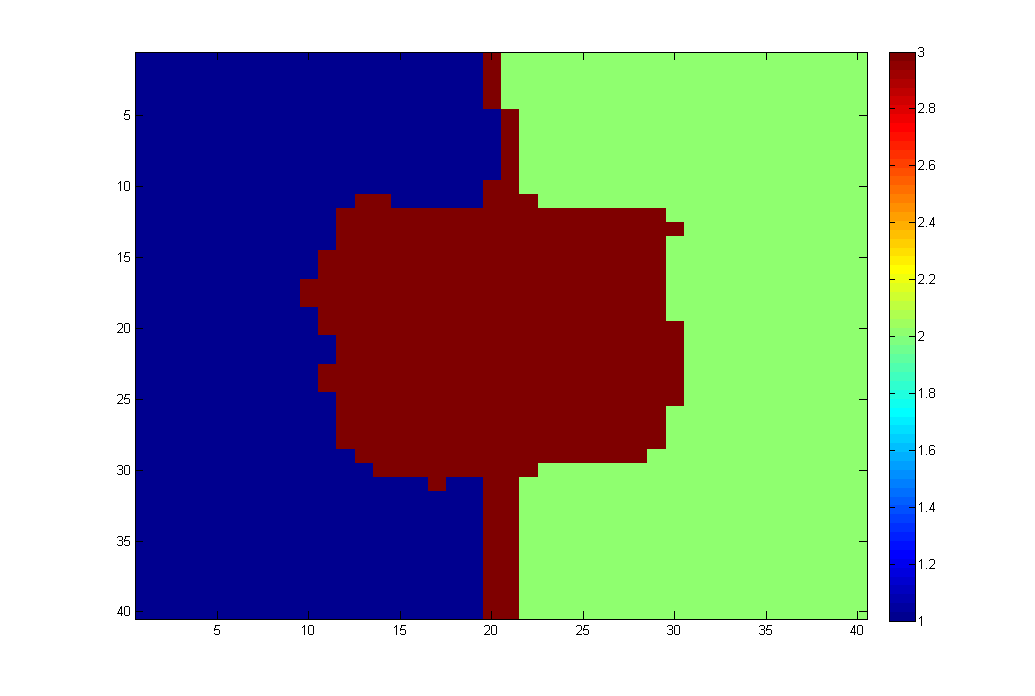}}
\end{minipage}%
\vspace{-0.1 in}
\caption{(a) True $\beta$ values for three regions; (b) Smooth residual effects (c) Y  values; (d)  dfbetas (e) Detected pattern of $\hat{\beta}$.}
\label{fig:rs}
\end{figure}

We first generate a map of spatial units from three distinct regions by letting $i=1, \cdots S$ and $i \in R^c$ with $c=1,2,3$. We assume that there are three distinct associations $\beta^c$ for the three different regions. Then, we simulate the covariates $x_i \sim N(\mu_x, \tau^2)$ and residuals $\epsilon \sim N(0,\sigma^2)$, for example, we let $\mu=5, \tau=2, \sigma=1$. We further apply Gaussian kernel to smooth the residuals to reflect the spatial dependency.  Then, the dependent variable follows $y_i=x_i \beta^c+\epsilon$. The input data are observations of ${x_i, y_i}$ for  $i=1, \cdots S$ and the region parcellation is unknown. We illustrate the data simulation and model fitting procedure in Figure \ref{fig:rs}.

We simulate 100 data sets at three of the noise levels ($\sigma^2=1$, $\sigma^2=10$, and $\sigma^2=100$). We apply the RAR method to analyze the simulated data sets and compare it to the CAR model and SAR model. 
 We evaluate the performance of different methods by using the criteria of the bias and 95\% CI coverage of $\beta$ across the 100 simulated data sets, and the results are summarized in Table 1. We note that though some smoothing effects are observed at the boundaries, the transitions are fairly well recaptured. We do not include the Bayesian CAR spatially varying coefficient model (e.g. \citealp{Wheeler08} ) because it yields different regression coefficient for each location $i$, and it is not available for the comparison of region level  $\beta$.  The results show that without region parcellation to account for the spatially varying coefficients, the EHA estimation could be vastly biased by using conventional spatial data analysis methods. In addition,    RAR model seems not to be affected by the noise levels. The RAR method can effectively and reliably detect the spatially varying regions and yield robust and close estimates of true $\beta^c$.

\begin{table}[h!]
\caption{Summaries of Estimates of $\hat{\beta}^c$ in the Simulation
Study}
\begin{center}
\begin{tabular} {lccccccc}
\toprule
\toprule
& \multicolumn{2}{c}{RAR} & \multicolumn{2}{c}{SAR} &  \multicolumn{2}{c}{CAR}\\
Parameters  & Mean (sd) &  CI coverage  &  Mean (sd) & CI coverage & Mean (sd) &CI coverage\\
\hline
 
 &&&$\sigma^2=1$ &&\\ 
$\beta^1$ (40)  & 39.00(3.22)   &  98\% &  4.19 (0.35)  & 4.1\%   &  3.35 (0.28) & 4.9\%\\
$\beta^2$ (-30) & -28.68 (1.93)& 99\% &   -   &  6.3\% & - & 5.6\%\\
$\beta^3$ (10) &  9.78 (0.35)& 99\% &  -  &  22.3\%  & - &19.3\%\\
\hline
 
 &&&$\sigma^2=10$ &&\\
$\beta^1$ (40)  & 39.13(2.89)   &  99\% &  2.07 (1.12)  & 1.5\%   &  2.23 (1.54) & 1.3\%\\
$\beta^2$ (-30) & -29.14 (1.02)& 99\% &   -   &  4.3\% & - & 2.7\%\\
$\beta^3$ (10) & 8.88 (0.35)& 98\% &  -  &  8.3\%  & - &5.9\%\\
\hline
 &&&$\sigma^2=100$ &&\\
$\beta^1$ (40)  & 39.27(3.14)   &  99\% &  1.98 (0.87)  & 0.9\%   & 0.8 (0.93) & 0.4\%\\
$\beta^2$ (-30) & -29.32 (1.87)& 99\% &   -   &  0.2\% & - &0.3\%\\
$\beta^3$ (10) &  8.77 (0.52)& 99\% &  -  &  2.5\%  & - &2.1\%\\
\bottomrule
\end{tabular}
\end{center}
\end{table}

\section{Data Example: air pollution and cardiovascular disease death rate association study}

\begin{figure}[ht!]
\centering \caption{Air pollution and mortality}
\includegraphics[scale=0.45]{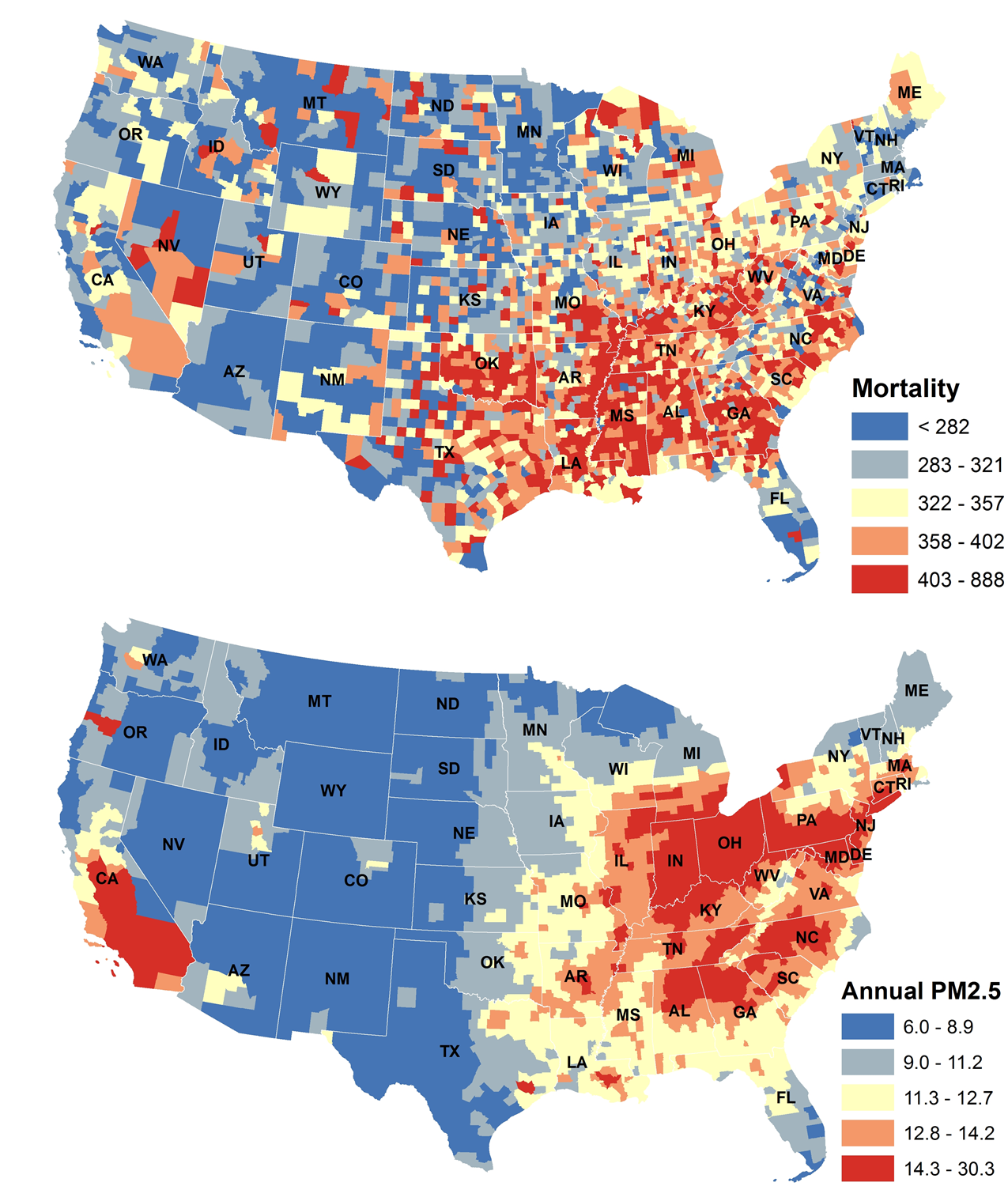}
\label{x}
\end{figure}

The 2010 annual circulatory mortality with ICD10 code I00-I99 and annual ambient fine particular matter (PM2.5) for 3109 counties in continental U.S was downloaded from CDC Wonder web portal. The annual mortality rate (per 100,000) was age-adjusted and the reference population was 2000 U.S population. The annual PM2.5 measurement was the average of daily PM2.5 based on 10km grid which were aggregated for each county. The measurement of PM2.5 in 10km grid was from US Environmental Protection Agency (EPA) Air Quality System (AQS) PM2.5 in-situ data and National Aeronautics and Space Administration (NASA) Moderate Resolution Imaging Spectroradiometer (MODIS) aerosol optical depth remotely sensed data. In this dataset, a threshold of 65 micrograms per cubic meter was set to (left) truncate the data to avoid invalid interpolation of grid PM2.5. We use the annual data set for 2001, because the population size and demographic information is more accurate by using 2000 U.S. census data. In figure \ref{x}, we illustrate the maps for annual age adjusted cardiovascular disease rate and air pollution (PM 2.5) level.


\begin{figure}[ht!]
\centering \caption{Results: spatially varying associations on RAR detected regions}
\includegraphics[scale=0.65]{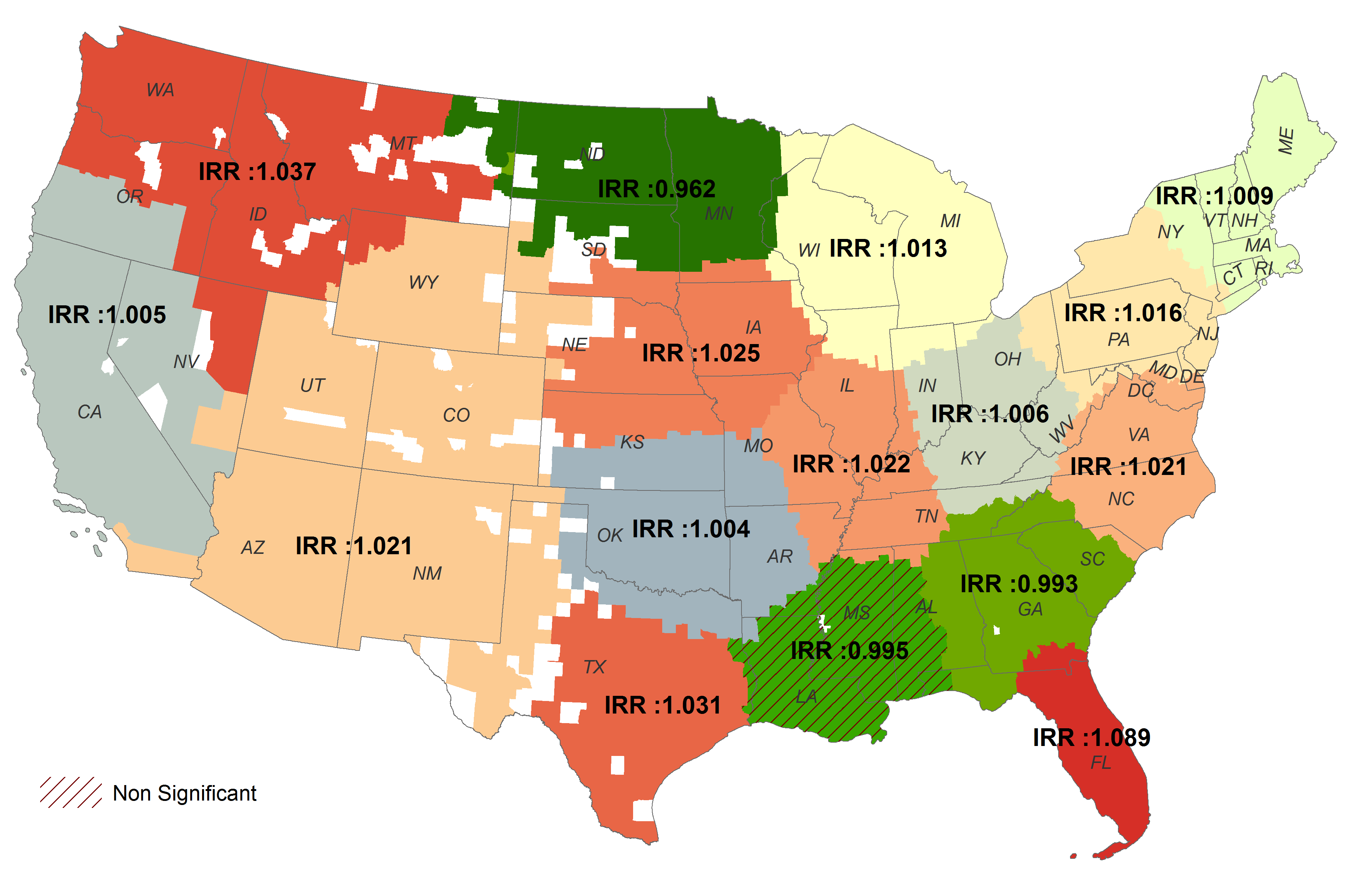}
\label{beta2001}
\end{figure}

We perform the RAR analysis on this data set by following the three steps described in section 2. We select the number of regions $K=16$ which minimizes BIC. We then incorporate  the identified region labels as categorical covariates  as well as the interaction between region labels and PM2.5 exposure levels. We examine the effects of detected regions (by introducing 15 dummy variables and 15 interaction variables) by using likelihood ratio test, both main effects (region) and interaction terms are significant with $p<0.001$. Figure \ref{beta2001} demonstrates the automatically detected regions and spatially varying associations between PM 2.5 and mortality rate at different regions (secondary parameters). The results reveal that the EHA are not coherent across the counties in the nation and RAR defines regions by breaking and rejoining the counties in different states based on similarity of EHA. The RAR defined map may also reveal potential confounders that affect health risk assessment. We note that the most significant positive EHA resides in northwest region and Florida: although the air pollution levels are not among the highest, the disease and exposure are highly positively associated. There are regions exhibiting negative EHA, for instance, the southeast region (part of GA, SC, and NC) and we further verify the association on the exposure and health outcome in enlarged Figure and interestingly by visual checking the exposure and health outcome are negatively associated. There could be potential co-founders such as medical facility accessibility and dietary behavior difference, etc. Our results reveal that region-wise EHA may vary across the nation (affected by local factors), which in part addresses the ecological fallacy (Simpson's Paradox). The RAR method could be used  as a tool to raise further research questions and to motivate new public health research investigation of the variation.

\begin{figure}[ht!]
\centering \caption{Enlarged exposure and mortality maps to illustrate the negative EHA in the southeast}
\includegraphics[scale=0.65]{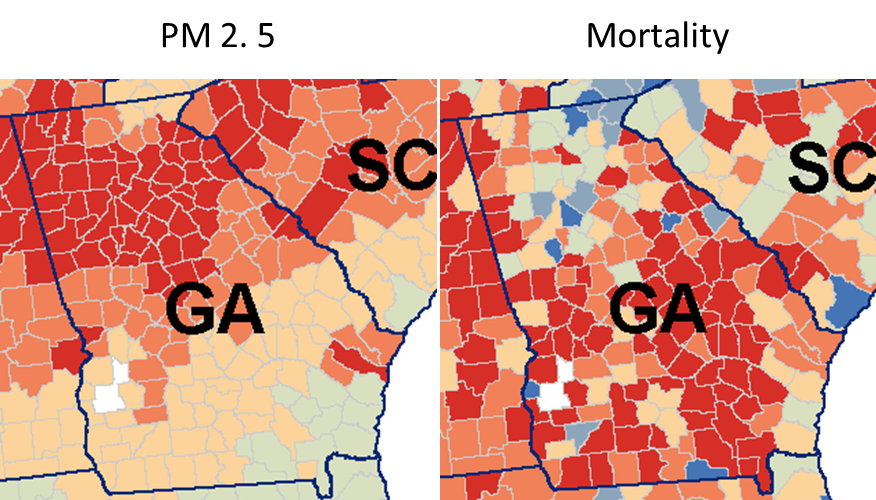}
\label{zoom}
\end{figure}



\section{Discussion}
 
The mapping of disease  and further building environment-health/disease association have  long been a key aspect of  public health research. However there has been challenge for statistical data analysis to yield spatially varying EHA at region level. Most previous methods employ random effect model by letting each spatial unit have a random slope and borrow power from neighbors, which is advantageous with regard to improving model fitting and model variance explanation. But, it is also beneficial to provide a map of regions that is defined by EHA similarity because it would allow us to directly draw statistical inferences at the region level as main effects may vary spatially(which is our major motivation to develop the RAR method). 

Rooted from image parcellation, the RAR framework aims to parcellate a large map into several contiguous regions with a two-fold goal: i) to define data-driven regions that reveals spatially varying EHA at region level; and ii) to utilize  the regions to fit a better regression model. Based on our simulation and example data analysis results, it seems that the performance of the RAR method is satisfactory. A CAR or SAR model could be applied on top of the RAR method, but the RAR could uniquely provide region-wise inferences.  The region-wise EHA inferences provide important guidance for public health decision making. For instance, although the exposure levels at California and Ohio River Valley are high, the EHA of these two regions are not among the highest associations. Thus, the RAR model could provide the informative geographic parcellations and inferences that neither exposure or health status data alone  can exhibit (e.g Figure 2).

Clustering and cluster analysis have been widely applied in spatial data analysis (\citealp{Waller15}). However, the most methods are limited to provide disjoint and contiguous regions with distinct disease exposure associations. The regions with different associations may suggest some unidentified confounders and other public health or geographic/environmental factors affecting population health.   
Although we demonstrate our model for cross sectional analysis, a longitudinal model could be extended straightforwardly because the DFBETAs could be computed for GEE or mixed effect model as well. The computational load of the RAR model is negligible and all of our simulation and data example calculation time is within a minute by using a PC with i7 CPU and 24G memory.

\section*{Acknowledgement }
Chen's research is supported in part by UMD Tier1A seed grant.

\section*{Appendix}
\noindent\textbf{Algorithm}:\newline 
Given an adjacency matrix $W_s$ for each subject $s$ and the number of classes $K$,
\begin{enumerate}
  \item[Step 1] Compute the degree matrix  $D_s$, where $D_s(i,i)=\sum_{j=1}^N W_s(i,j)$.
  \item[Step 2] Find the eigen-solutions $[V_s,L_s]$ of  $D_s^{-\frac{1}{2}}W_sD_s^{-\frac{1}{2}}$ ,  i.e., solving   $D_s^{-\frac{1}{2}}W_sD_s^{-\frac{1}{2}}V_s=V_sL_s$  and  $V_sV_s=I_N$. Then compute $Z_s=D_s^{-\frac{1}{2}}V_s$.
  \item[Step 3] Normalize $Z_s$ by  setting   $X_s={\mathrm{diag}}^{-1}[{\mathrm{diag}}(Z_s Z_s^T)] Z_s$,  where operation $\mathrm{diag}(A)$ extracts the diagonal elements of matrix $A$ as a vector; and $\mathrm{diag}^{-1}(a)$ creates a matrix with diagonal elements equal to $a$ and off-diagonal being zeros.  
  \item[Step 4] Set the convergence criterion parameter $\rho^*=0$,  and initialize a $K\times K$ matrix $R_s$ by the following steps: denote by $R_s^k$ the $k$th column of $R_s$ for $k = 1,\ldots, K$. 
 Set $R^1_s = [X_s(i,1),\ldots, X_s(i,K)]^{\mathrm T}$, where $i$ is randomly selected from $\{1,\ldots, N\}$. We denote the first column of $R_s$ as $R_s^1$ and the $k_{th}$ column as $R_s^k$.
   Then update the rest of the columns by following.\newline
{ For $k = 2,\ldots, K$, iteratively update
 $R^k_s = [X_s(i_k,1),\ldots, X_s(i_k,K)]^{\mathrm T}$ where 
 $$i_k = \argmin_{i\in\{1,\ldots,N\}}c_{k-1}(i), \mbox{ and }  c_{k-1} =  \sum_{l=1}^{k-1} |X_s R_s^{l}|.$$

}

 \item[Step 5] Minimize the objective function:
 $\sum_s||Y-X_sR_s||^2=||Y-\widehat{XR}||^2$.
 \newline 
 where $||\cdot ||$ stands for Frobenius norm; and $\widehat{X R}=\sum_s t_s X_sR_s$ with $$t_s=\frac{1/||X_sR_s-X cR c||^{2}}{\sum_s 1/||X_sR_s-X cR c||^{2}}.$$
 The term $XcRc$ is the centroid  of $X_sR_s$   which minimizes $\sum_{s'\neq s}||X_sR_s-X_{s'}R_{s'}||^2$  with respect to $X_{s'}R_{s'}$.  

Then $Y(i,l)=1$, where $l=\argmax_{k\in \{1,...,K\}}\widehat{XR}(i,k), \  \: i \in \{1,...,N\} \ and \ l\in \{1,...,K\}$.
\item[Step 6] Conduct singular value decomposition on the matrix $Y^TX_s$\newline
$Y^TX_s=U_s\Omega_s{V_s}^T$\newline
$\rho=\sum_s tr(\Omega_s)$\newline
If $|\rho-\rho^*| < \text{pre-assigned error limit then output} \: Y$,\newline
else, update $R_s=V_sU_s^T$.
\item[Step 7] Go to Step 5.
\end{enumerate}

\bibliographystyle{plainnat}

\end{document}